# Semi-Instrumental Variables: A Test for Instrument Admissibility

**Tianjiao Chu**

tchu@andrew.cmu.edu

**Richard Scheines**

Philosophy Department
Carnegie Mellon University
Pittsburgh, PA 15213
scheines@cmu.edu

**Peter Spirtes**

ps7z@andrew.cmu.edu

## Abstract

In a causal graphical model, an instrument for a variable $X$ and its effect $Y$ is a random variable that is a cause of $X$ and independent of all the causes of $Y$ except $X$ (Pearl 1995). For continuous variables, instrumental variables can be used to estimate how the distribution of an effect will respond to a manipulation of its causes, even in the presence of unmeasured common causes (confounders). In typical instrumental variable estimation, instruments are chosen based on domain knowledge. There is currently no statistical test for validating a continuous variable as an instrument. In this paper, we introduce the concept of semi-instrument, which generalizes the concept of instrument: each instrument is a semi-instrument, but the converse does not hold. We show that in the framework of additive models, under certain conditions, we can test whether a variable is semi-instrumental. Moreover, adding some distribution assumptions, we can test whether two semi-instruments are instrumental. We give algorithms to test whether a variable is semi-instrumental, and whether two semi-instruments are both instrumental. These algorithms can be used to test the experts' choice of instruments, or to identify instruments automatically.

## 1 INTRODUCTION

One of the major advantages of knowing the causal relations among the variables we are interested in is that we can use the causal structure to predict the effects of interventions. For example, if we know that a random variable $X$ is the cause of $Y$, then intervening on $X$ will change $Y$, but intervening on $Y$ should not change $X$.[1] By regressing $Y$ on $X$, we can estimate the effect of the intervention of $X$ on $Y$, up to a constant, if no unmeasured common cause of $X$ and $Y$ exists.

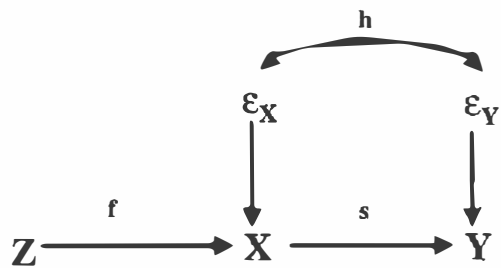

Figure 1: One Instrumental Variable

However, this advantage becomes problematic if we know, or even suspect, that there are some unmeasured common causes of $X$ and $Y$, in which case we cannot estimate the direct effect of $X$ on $Y$.

Consider the the causal structure illustrated in figure 1. Suppose that $Z$ is not observable for the moment. Among $X$, $Y$, $\epsilon_X$, and $\epsilon_Y$, only $X$ and $Y$ are observed variables. The functional relations among them are:

---

[1] Here by intervention we mean the manipulation of the value of one random variable, with the assumption that this manipulation can affect other variables only through the manipulated variable. We also assume no feedback, i.e., no directed cycles, and no reversible causal relations, in a causal graph. See Spirtes et al (2001)

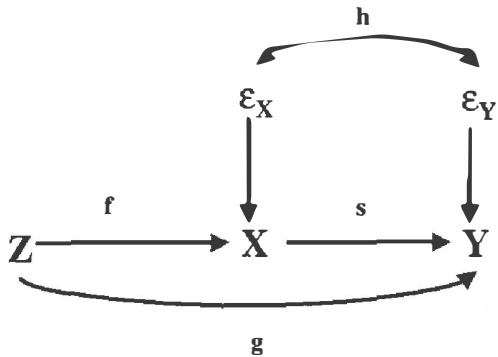

Figure 2: One Common Cause

$$X = f(Z) + \epsilon_X \quad (1)$$
$$Y = s(X) + \epsilon_Y \quad (2)$$
$$E[\epsilon_Y|\epsilon_X] = h(\epsilon_X) \quad (3)$$

Note that $\epsilon_Y$ and $X$ are dependent, that is, $E[\epsilon_Y|X]$ will be a non-constant function of $X$, say, $\lambda(X)$. The regression of $Y$ on $X$ will give:

$E[Y|X] = s(X) + E[\epsilon_Y|X] = s(X) + \lambda(X)$

Because we have no way to estimate $\lambda(X)$, we also have no way to identify $s(X)$, even up to a constant.

However, with the help of variable $Z$, we can estimate $\epsilon_X$ by regressing $X$ on $Z$ to get an estimate of $\epsilon_X = X - E[X|Z]$. Then, we can regress $Y$ on $X$ and $\epsilon_X$ to get:

$E[Y|X, \epsilon_X] = s(X) + E[\epsilon_Y|\epsilon_X, X]$

$= s(X) + E[\epsilon_Y|\epsilon_X] = s(X) + h(\epsilon_X)$. [2]

An additive regression method will give an estimate of $s(X)$ and $h(\epsilon_X)$ simultaneously.[3]

Here $Z$ is called an instrumental variable for $X$ and $Y$, by which we mean that $Z$ is a direct cause of $X$, and $Z$ is independent of all other causes of $Y$ except $X$. (Note that $X$ must be a direct cause of $Y$.)

---

[2] Here we use the fact that $X$ and $\epsilon_Y$ are independent conditional on $\epsilon_X$, which is implied by the causal graph in figure 1.

[3] For the proof of the identifiability of $s(X)$, up to a constant, see Newey et al (1999).

## 2 PRIOR WORK

There has been intensive study of the use of instruments in the econometric literature. Much work has focused on the study of the efficiency of an instrument (Nelson et al 1990, Shea 1997, and Zivot et al 1998). There are studies of whether a variable $Z_1$ is an instrument for $X$ and $Y$, provided we already have an instrument $Z_2$ for $X$ and $Y$ (Wu 1973, Newey 1985, and Magdalinos 1994), but how to find the first instrument is still an open problem. There is no test available to find out whether a continuous variable is an instrument, without knowing that some other variable is an instrument. [4]

In practice, people usually resort to domain knowledge to determine whether a variable is an instrument. As some studies show, moreover, it is often very difficult to find instruments, and a wrong choice may significantly affect the final result (Bartels 1991, Heckman 1995, and Bound et al 1995). Therefore, even in the case where we do have some domain knowledge to help us identify instruments, some kind of testing procedure is still useful in that it can serve as a double check. It is the goal of this paper to find out whether, under certain reasonable conditions, we can test whether a variable is an instrument.

## 3 APPROACH

Within the framework of linear models, the fact that a variable is an instrument imposes no constraint on the joint marginal distribution of the observed variables. Consider the causal structure illustrated in figure 1, where only $Z$, $X$, and $Y$ are observable. Assume for now that all the functional relations are linear. In particular, we assume that:

$$X = fZ + \epsilon_X \quad (4)$$
$$Y = sX + \epsilon_Y \quad (5)$$

where $f$ and $s$ are non-zero constants.

Now we construct another model based on the causal structure illustrated in figure 2. Let $g' = c \neq 0$, $s' = s - \dfrac{g'}{f}$, and $\epsilon'_Y = \epsilon_Y + \dfrac{g'}{f}\epsilon_X$. Let $Z' = Z, X' = X$, and $Y' = g'Z' + s'X' + \epsilon'_Y$. Clearly $Y' = Y$, hence $(Z, X, Y)$ has the same distribution as $(Z', X', Y')$. However, we notice that in the first case, $Z$ is an instrument for $X$ and $Y$, while in the second case, $Z'$ is not an instrument, but a common cause of $X'$ and $Y'$. Actually, we can set $g'$ to be any value, and adjust $s'$ and $\epsilon'_Y$ accordingly so that $X' = X$ and $Y' = Y$. This means that, a joint distribution implied by a model where $Z$ is an instrument could also be implied by uncountably many other models where $Z$ is not an instrument. Based on the joint distribution, we cannot tell whether $Z$ is an instrument.

---

[4] Pearl (1995) shows that, for discrete data, $Z$ being an instrument for $X$ and $Y$ imposes some constraint, which is called by him as *instrumental inequality*, on the joint distribution of $Z$, $X$, and $Y$. This inequality is not sufficient for $Z$ to be an instrument, and, as pointed in this paper, it cannot be extended to the continuous models. This paper does not give an instrument testing procedure for discrete data.

The above analysis suggests two possible approaches for instrument testing. First, we may extend the space of linear structural models to a larger space. The idea is that the space of linear structural models already imposes strong constraints on each variable so that being an instrument does not impose any extra constraints. However, in a larger space, it is possible that being an instrument does imply some conditions not necessarily satisfied by all models in that space. Another approach is to get more candidates for instrumental variables, and see whether the instrument assumption imposes some constraint on the joint distribution of these candidate variables.

In this section we shall try a combination of both approaches. That is, we shall propose a two-stage procedure. Suppose we are given a set of candidate instrument variables. In the first stage of the test, we shall exclude some, but not necessarily all, non-instruments. In the second stage of the test, we shall determine whether all the remaining ones are instruments, provided we still have at least two candidates left.

### 3.1 SEMI-INSTRUMENTAL VARIABLES FOR THE ADDITIVE NONPARAMETRIC MODELS

Here by an additive nonparametric model we mean that, for each endogenous variable, its expectation given its parents is a linear combination of univariate functions of its parents, plus some error term with unknown distribution. We further assume that all exogenous variables are independent, and that they are independent of all the error terms. We do allow dependence between the error terms of $X$ and $Y$. Later in this paper we shall use additive model and additive nonparametric model interchangeably.

Figure 1 and figure 2 give two very simple additive models. We can take them as two alternative hypotheses about the underlying model that generates a sample with observable variables $Z$, $X$, and $Y$. The problem of testing whether $Z$ is an instrument for $X$ and $Y$ is equivalent to the problem of determining which figure represents the correct model.

Note that a linear model is a special case of the additive model. Thus, given that we cannot determine whether a random variable is an instrument in a linear model, we could not, in general, determine whether it is an instrument in an additive model. Therefore, we should look for some other conditions, preferably a little bit weaker than those required by an instrument, so that a random variable that satisfies these conditions can be identified in an additive model.

One such candidate set of conditions is what we call the semi-instrumental conditions. A random variable $Z$ is a *semi-instrument* for $X$ and $Y$ if the following conditions are satisfied:

1. *The joint distribution of $Z$, $X$, and $Y$ can be represented by an additive model $M$ consisting only of $Z$, $X$, and $Y$ as the observed variables;*

2. *$Z$ is an exogenous variable in $M$;*

3. *$Z$ is the cause of $X$, and $X$ is a cause of $Y$ in $M$;*

4. *If $Z$ is also a cause of $Y$, then the direct effect of $Z$ on $Y$ is a linear function of the direct effect of $Z$ on $X$.*

Note that if a random variable $Z$ is a semi-instrument for $X$ and $Y$, then the direct effect of $Z$ on $Y$, say, $g(Z)$, is a linear function of the direct effect of $Z$ on $X$, say, $f(Z)$. That is, there is a pair of real numbers $(a, b)$ such that $g(Z) = af(Z) + b$. We shall call $a$ the *linear coefficient of the semi-instrument $Z$*.

It is easy to see that the semi-instrumental assumption is weaker than the instrumental assumption: All instruments are semi-instruments (with a linear coefficient 0), but not all semi-instruments are instruments. Moreover, in general, in a linear model, an exogenous $Z$ that is a common cause of $X$ and $Y$, which could not be an instrument, is a semi-instrument for $X$ and $Y$, because both its effect on $X$, i.e., $f$, and its effect on $Y$, i.e., $g$, are linear functions. Therefore, by extending the space of linear models to the space of additive models, we find that the instrument assumption does impose some constraints on the possible kinds of effects of $Z$ on $Y$: Only models where the effect of $Z$ on $Y$ is a linear function of the effect of $Z$ on $X$ are compatible with the distribution implied by a model where $Z$ is an instrument.

To test whether a random variable $Z$ is a semi-instrument for $X$ and $Y$, theoretically we should check whether all the four conditions are satisfied. However, it turns out that not all these four conditions are testable. For example, the second condition, i.e., $Z$ is an exogenous variable, cannot be tested if we only have the joint distribution of $X$, $Y$, and $Z$. From the joint distribution only, there is no way to tell whether $Z$ and the error term associated with $X$, say, $\epsilon_X$, are dependent.

On the other hand, we do have many cases where it is reasonable to assume that the first three conditions are satisfied. For example, it is typical that when testing whether $Z$ is an instrument for $X$ and $Y$, we are dealing with an additive model described by conditions 1 to 4, and our question is whether $Z$ is also a direct cause of $Y$.

Assuming that the first three conditions are satisfied, under certain smoothness conditions, to test whether the fourth condition is also satisfied is equivalent to testing whether $E[Y|X,\epsilon_X]$ is a linear combination of a univariate function of $X$ and a univariate function of $\epsilon_X$:

**Theorem 1** *Consider the additive model given by figure 2. Suppose that $(X,\epsilon_X)$ has a joint density, and that all the functions, i.e., $f$, $g$, $s$, and $h$, are differentiable.* [5] *Then $E[Y|X,\epsilon_X]$ is a linear combination of a univariate function of $X$ and a univariate function of $\epsilon_X$, and $Var(Y|Z,\epsilon_X) = Var(Y|E[X|Z],\epsilon_X)$, if and only if $g(Z) = af(Z) + b$ for some pair of real numbers $(a,b)$.*

For the proof, see the appendix.

The above theorem suggests an algorithm, which we will call the *semi-instrument testing algorithm*, to test whether $Z$ is a semi-instrument for a sample $S$ generated from an additive model given by figure 2. This test has two steps, the first step is the additivity test for the null hypothesis that $E[Y|X,\epsilon_X]$ is a linear combination of a univariate function of $X$ and a univariate function of $\epsilon_X$:

1. Regress $X$ on $Z$ to estimate $\epsilon_X = X - E[X|Z]$

2. Regress $Y$ on $X$ and $Z$ with a surface smoother, and score the fit of this model with some score function that penalizes model complexity.

3. Regress $Y$ on $X$ and $\epsilon_X$ with an additive smoother, and score the fit of this model with some score function that penalizes model complexity.

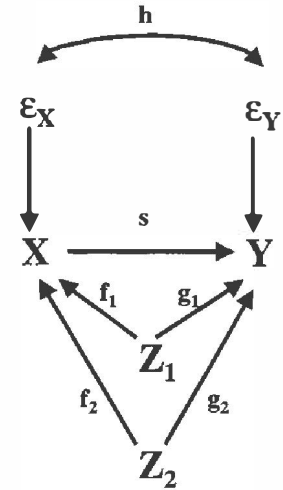

Figure 3: Two Common Causes

4. If the additive model has a better score, accept the null hypothesis. Otherwise, reject it.

The second step is the measurability test for the null hypothesis that $Var(Y|Z,\epsilon_X) = Var(Y|E[X|Z],\epsilon_X)$:

1. Regress $X$ on $Z$ to estimate $E[X|Z]$ and $\epsilon_X$,

2. Regress $Y$ on $Z$ and $\epsilon_X$ with a surface smoother, and let $R_A$ be the sum of the residuals.

3. Regress $Y$ on $E[X|Z]$ and $\epsilon_X$ with a surface smoother, and let $R_N$ be the sum of the residuals.

4. If the regression of $Y$ on $Z$ and $\epsilon_X$ has a smaller sum of residuals, i.e., $R_A < R_N$, reject the null hypothesis. Otherwise, accept it. [6]

### 3.2 TWO SEMI-INSTRUMENTS

If the test for whether a random variable, say, $Z_1$, is a semi-instrument for $X$ and $Y$ gives a negative result, there is not much left to do with $Z_1$. However, if the test says that $Z_1$ is a semi-instrument, we will face another problem: Is $Z_1$ an instrument?

We have pointed out that this question cannot be answered if we only have the joint distribution of $Z_1$, $X$, and $Y$. However, from the Bayesian point of view, with some further assumption, if there is a second semi-instrument, say, $Z_2$, we might be able to determine whether $Z_1$ and $Z_2$ are both instruments. (If not both of them are instrument, we would not be able to tell whether both are non-instrumental, or only one is non-instrumental.) The following theorem gives the condition when two semi-instruments are both instruments almost surely.

---

[5] These two conditions are much stronger than what we need. Actually, we only need to assume the boundary of the support of the $(X,\epsilon_X)$ has no positive probability, (see proof of Theorem 2.3 in Newey et al (1999)), and that all the functions are absolutely continuous. We choose these two conditions because they are more intuitive.

[6] Here we might also want to do a bootstrap estimation of the distribution of $R_N - R_A$. If we adopt this approach, we could generate bootstrap samples by adding permutations of the residuals obtained by regressing $Y$ on $E[X|D]$ and $\epsilon_X$ to the fitted values of $Y$ obtained from the same regression.

**Theorem 2** *Let $Z_1$ and $Z_2$ be two independent random variables that are both semi-instruments for $X$ and $Y$. Let $a_1$ and $a_2$ be the linear coefficients of $Z_1$ and $Z_2$ respectively. Suppose $a_1$ and $a_2$ are independent, and each has a distribution function that has one and only one point of discontinuity, 0.* [7] *If $Z_1$ and $Z_2$ have the same linear coefficients, then with probability 1, $Z_1$ and $Z_2$ are both instruments.*

For the proof, see the appendix.

Assume that the sample $S$ was generated from the causal structure illustrated in figure 3, and that both $Z_1$ and $Z_2$ are semi-instruments for $X$ and $Y$. We can use the following algorithm, called the *double instruments testing algorithm*, to test whether $Z_1$ and $Z_2$ have the same linear coefficient: [8]

1. Create a new variable $Z = f_1(Z_1) + f_2(Z_2)$, where $f_1(Z_1) = E[X|Z_1]$, $f_2(Z_2) = E[X|Z_2]$.

2. Test whether $Z$ is a semi-instrument.

This algorithm is based on the following observation:

Assume that $g_1$ and $g_2$ are differentiable, and that $Z_1$ and $Z_2$ have a joint density. Then $g_1(Z_1) + g_2(Z_2) = a(f_1(Z_1) + f_2(Z_2)) + b$ for some $(a, b)$ iff $g_1(Z_1) = af_1(Z_1) + b_1$ and $g_2(Z_2) = af_2(Z_2) + b_2$ for some $b_1 + b_2 = b$. [9]

## 4 SIMULATION

We have done some simulation studies to estimate the performance of the two algorithms for semi-instrument testing and double instrument testing. It turns out that, in order for the semi-instrument testing to work, we need to find a better additive regression method that can handle dependent predictors, which seems currently not available. However, the double-instruments testing algorithm does work for a subset of additive models: the models where the influence of $X$ on $Y$ is linear. This subset includes a very important class of models: the linear models. [10]

### 4.1 SEMI-INSTRUMENT TESTING

The semi-instrument testing algorithm requires a surface smoother and an additive smoother. We use the Splus functions `loess` as the surface smoother, and `gam` as the additive smoother. The `gam` function implements the back-fitting algorithm proposed in Hastie et al (1990). The `loess` function is an implementation of the local polynomial regression. We use BIC score function to score the fitted models returned by `gam` and `loess` respectively.

We generate 6 samples from the following 2 models:

$X = Z^2 + \epsilon_X$

$Y_i = X^2 + c_i Z^3 + \epsilon_Y$

where $E[\epsilon_Y|\epsilon_X] = \epsilon_X^2$, $c_1 = 0$, and $c_2 = 1$.

These two models share the same $Z$, $\epsilon_X$, $X$, and $\epsilon_Y$, differ in the effect of $Z$ on $Y$, and hence differ in $Y$. In the first model, $Z$ is an instrument, hence a semi-instrument. In the second model, $Z$ is not a semi-instrument. [11] For each model, we generated 3 samples with sizes 200, 1000, and 5000 respectively.

Table 1: `gam` and `loess` models comparison

| size | model | gam BIC | loess BIC |
|---|---|---|---|
| 200 | 1 | 486.8 | 239.4 |
| 1000 | 1 | 2530.3 | 1041.8 |
| 5000 | 1 | 12438.3 | 5053.4 |
| 200 | 2 | 350.9 | 238.8 |
| 1000 | 2 | 1404.6 | 1041.8 |
| 5000 | 2 | 6670.7 | 5053.4 |

From the above data, we can see that the `gam` model is always much worse than the loess model, no matter whether $Z$ is a semi-instrument or not. This implies that no matter whether the null hypothesis is true, i.e., $Z$ is a semi-instrument, the test procedure will always reject it! The most plausible explanation of this phenomenon is that because of the dependence between $\epsilon_X$ and $X$, the performance of gam is significantly worse than that of loess. It seems that the back-fitting algorithm often gets trapped in a plateau, and in some cases fails to converge.

---

[7] Note that here by imposing a distribution on $a_1$ and $a_2$, which are actually parameters of our models, we have adopted a Bayesian perspective. Also, the conditions for the distribution are stronger than required. What we really need is to ensure that the $P(a_1 = a_2) = P(a_1 = a_2 = 0) > 0$. That is, we want to assume that it is possible that $a_1 = a_2$, and if $a_1 = a_2$, it is almost sure that $a_1 = a_2 = 0$, which means that both $Z_1$ and $Z_2$ are instruments.

[8] Note that from the classical point of view, this algorithm can be used to reject the null hypothesis that $Z_1$ and $Z_2$ are both instruments in the cases where $Z_1$ and $Z_2$ have *different* linear coefficients. However, to make it a testing algorithm for double instruments, a certain Bayesian assumption about the prior distributions of the linear coefficients of $Z_1$ and $Z_2$, like the one proposed in Theorem 2, is required.

[9] The proof of this observation is similar to that of Theorem 1.

[10] The second algorithm works for the models where the influence of $X$ on $Y$ is linear because in this case we can modify the algorithm so that we do not need to apply additive regression method to models with dependent predictors. For a detailed discussion, see section 4.2.

[11] The distribution of these variables are: $Z$ is uniform between 0 and 5. There is also a latent variable $T$ that is uniform between 0 and 2. $\epsilon_X$ is the sum of $T$ and a normal noise with standard deviation 0.5, $\epsilon_Y$ is the sum of $T^2$ and a normal noise with standard deviation 0.5.

### 4.2 DOUBLE INSTRUMENTS TESTING

Despite the lack of good additive regression method, with some further conditions, we could still make the double instruments testing algorithm work. Consider the model given by figure 3. If besides assuming $Z_1$ and $Z_2$ are semi-instruments, we further assume that $s(X) = cX + d$, i.e., the direct effect of $X$ on $Y$ is a linear function in $X$, then we will be able to test whether $Z_1$ and $Z_2$ have the same linear coefficients.

Let $g_2(Z_1) = a_1 f_1(Z_1) + b_1$, $g_2(Z_2) = a_2 f_2(Z_2) + b_2$, and $d_0 = b_1 + b_2 + d$ we have:

$$Y = cX + d + \epsilon_Y + g_1(Z_1) + g_2(Z_2)$$
$$= (c + a_1)f_1(Z_1) + (c + a_2)f_2(Z_2) + c\epsilon_X + \epsilon_Y + d_0.$$

That is, $Y$ is additive in $Z_1$, $Z_2$, where $Z_1$, $Z_2$ are independent of each other, and jointly independent of $\epsilon_X$ an $\epsilon_Y$.

If $Z_1$ and $Z_2$ have the same linear coefficient, i.e., $a_1 = a_2$, [12] we further have:

$$Y = (c + a_1)[f_1(Z_1) + f_2(Z_2)] + c\epsilon_X + \epsilon_Y + d_0.$$

Let $Z = f_1(Z_1) + f_2(Z_2)$, it is easy to see that:

$\text{Var}(Y|Z_1, Z_2) = \text{Var}(c\epsilon_X + \epsilon_Y)$

$\leq \text{Var}(c\epsilon_X + \epsilon_Y) + \text{Var}(a_1 f_1(Z_1) + a_2 f_2(Z_2)|Z)$

$= \text{Var}(Y|Z)$

with the equality holds only when $a_1 = a_2$.

The following algorithm, which is called the *linear double instruments testing algorithm*, compares the mean square error of regressing $Y$ on $Z_1$ and $Z_2$, with the mean square error of regressing $Y$ on $Z = \mathrm{E}[X|Z_1, Z_2]$. It can be used to test whether the two semi-instruments $Z_1$ and $Z_2$ have the same linear coefficients, assuming the direct effect of $X$ on $Y$ is linear in $X$:

1. Use additive regression to regress $X$ on $Z_1$ and $Z_2$ to get $\mathrm{E}[X|Z_1, Z_2]$.

2. Let $Z = \mathrm{E}[X|Z_1, Z_2]$. Regress $Y$ on $Z$ using linear regression, and compute the BIC score $BIC_l$.

3. Regress $Y$ on $Z_1$ and $Z_2$ using additive nonparametric regression, and compute the BIC score $BIC_a$.

4. $Z_1$ and $Z_2$ have the same linear coefficients if $BIC_l < BIC_a$.

To test the above algorithm, we generated 12 samples from the following 3 models:

$$X = Z_1^2 + Z_2^2 + \epsilon_X$$
$$Y_i = X + c_i Z_2^2 + \epsilon_Y$$

where $\mathrm{E}[\epsilon_Y|\epsilon_X] = \epsilon_X^2$, and $c_1 = 0, c_2 = 0.2$, and $c_3 = 1$.

These three models share the same $Z_1$, $Z_2$, $\epsilon_X$, $X$, and $\epsilon_Y$, but differ in the linear coefficients of $Z_1$ and $Z_2$, and hence differ in $Y$. In the first model, $Z_1$ and $Z_2$ have the same linear coefficients. In the second model, there is a small difference in the linear coefficients. In the third model, the difference is significant. [13] For each model, we generated 4 samples with sizes 50, 100, 200, and 500 respectively.

Table 2: Values of $BIC_a - BIC_l$ for the 12 samples

| sample size | $c_1 = 0$ | $c_2 = 0.2$ | $c_3 = 1$ |
|---|---|---|---|
| 50 | 11.35 | 8.79 | -95.57 |
| 100 | 14.23 | 1.60 | -333.81 |
| 200 | 13.07 | -2.53 | -419.26 |
| 500 | 19.10 | -35.56 | -1233.05 |

The entries of the above table are the values of $BIC_l - BIC_a$ for each sample. A positive value means that the null hypothesis is accepted for that sample. From the table, we can see that to detect the significant difference between the linear coefficients of $Z_1$ and $Z_2$, a sample of size 50 is sufficient. But when the difference is small, we need a sample size of 200.

## 5 DISCUSSION AND FUTURE WORK

### 5.1 THREE ASSUMPTIONS

The semi-instrument testing algorithm assumes that the first three semi-instrumental conditions are sat-isfied. While in general we cannot test whether a random variable satisfies all the first three semi-instrumental conditions, it is interesting to know whether we can test for one of them, especially the second semi-instrumental condition: $Z$ is an exogenous variable. The answer is: in principle, this assumption can be tested by the method of instrument, if we have an instrument for $Z$ and $X$. However, it is easy to see that this will lead to an infinite regression.

---

[12] Note that by Theorem 2, under certain distribution assumptions, $a_1 = a_2$ implies that $a_1 = a_2 = 0$ w.p.1, i.e., both $Z_1$ and $Z_2$ are instruments.

[13] The distribution of these variables are: $Z_1$ and $Z_2$ both are uniform between 0 and 4. There is also a latent variable $T$ that is uniform between 0 and 2. $\epsilon_X$ is the sum of $T$ and a normal noise with standard deviation 0.5, $\epsilon_Y$ is the sum of $T^2$ and a normal noise with standard deviation 0.5.

Another assumption key to the double instruments testing algorithm is: The prior probability that $Z$ is instrumental is positive, while the prior probability is 0 for a semi-instrument to have linear coefficient $a$ if $a \neq 0$. This raises a question: Why does the value 0 have a special status in the range of possible values of the linear coefficients of a semi-instrument? Here we want to give an argument for the plausibility of this assumption: If we take the set of possible causal structures among $X, Y, Z_1$ and $Z_2$ as a discrete sample space, it is reasonable to assign a positive prior probability to one element in the space, i.e., the structure where both $Z_1$ and $Z_2$ are instruments, which means that both $Z_1$ and $Z_2$ have linear coefficients 0. On the other hand, if a semi-instrument is not an instrument, there is no specific reason to believe that its linear coefficient should take any specific non-zero value.

We make a third assumption in an effort to modify the double instruments testing algorithm so that it has sufficient power: We assume that the direct effect of $X$ on $Y$ is a linear function of $X$. We notice that this is a rather strong assumption for an additive model. Moreover, because currently we do not have a suitable additive regression method for the semi-instrument testing, we also have to assume, without any testing, that $Z_1$ and $Z_2$ are semi-instruments. Nevertheless, the modified double instruments testing algorithm is general enough to provide a double instruments test for linear models.

## 5.2 FUTURE WORK

To make the semi-instrument testing powerful, we will continue to look for some additive regression method that is suitable for the case where the predictors are dependent. [14] Alternatively, we may also try to find some new ways of testing semi-instruments where the problem of the dependence of the predictors will not significantly affect the test results.

---

[14]Tom Minka suggested that by letting the back-fitting algorithm run for a sufficient number of iterations, it will eventually return a good fitted model, or we can use least squares to the the best fitted model directly (without iterations). We have yet to test whether this will work.

### Acknowledgments

This paper was supported by NSF grant DMS-9873442.

## Appendix

**Proof of Theorem 1.**

Because $(X, \epsilon_X)$ has a joint density, and $f, g, s$, and $h$ are differentiable, it immediately follows that $E[Y|X = x, \epsilon_X = u]$ is differentiable with respect to $X$ and $\epsilon_X$ with probability one.

Also note that $E[Y|X = x, \epsilon_X = u]$

$= E[s(X) + g(Z)) + \epsilon_Y | X = x, \epsilon_X = u]$

$= s(x) + E[g(Z)|f(Z) = x - u] + h(u),$

because conditional on $f(Z) = X - \epsilon_X = x - u$, $g(Z)$ is independent of $X = x$ and $\epsilon_X = u$

It is easy to see that if $g(Z) = af(Z) + b$, i.e., the direct effect of $Z$ on $Y$ is a linear function of the direct effect of $Z$ on $X$, then with probability one, $E[Y|X, \epsilon_X] = s(X) + aX + h(\epsilon_X) - a\epsilon_X + b$, i.e., $E[Y|X, \epsilon_X]$ is a linear combination of a univariate function of $X$ and a univariate function of $\epsilon_X$. Moreover, we have:

$\text{Var}(Y|E[X|Z], \epsilon_X)$

$= \text{Var}(s(f(Z) + \epsilon_X) + g(Z) + \epsilon_Y | f(Z), \epsilon_X)$

$= \text{Var}(\epsilon_Y | f(Z), \epsilon_X) = \text{Var}(\epsilon_Y | \epsilon_X)$

$= \text{Var}(\epsilon_Y | Z, \epsilon_X) = \text{Var}(Y | Z, \epsilon_X)$

To show the converse, suppose:

$E[Y|X = x, \epsilon_X = u] = s_1(x) + h_1(u).$

Let $g_0(x - u) = E[g(Z)|f(Z) = x - u]$, we have:

$\dfrac{\partial g_0(x - u)}{\partial x} = g_0'(x - u) = \dfrac{d}{dx}(s_1(x) - s(x))$

$\implies g_0'(x - u)$ is constant in $u$

$\implies g_0'$ is a constant

$\implies g_0$ is a linear function

Note that the assumption that $\text{Var}(Y|E[X|Z], \epsilon_X) = \text{Var}(Y|Z, \epsilon_X)$ implies that $\text{Var}(g(Z)|f(Z)) = 0$, which again implies that $g(Z) = E[g(Z)|f(Z)]$ w.p.1. Therefore, we have:

$g(Z) = af(Z) + b$, where $a, b$ are constants.

**Proof of Theorem 2.**

Let $L_1$ be the linear coefficient of $Z_1$, $L_2$ the linear coefficient of $Z_2$, and $\mu_{L_1}$ the distribution of $L_1$. Then:

$P(L_1 = L_2 = 0 | L_1 = L_2) = \dfrac{P(L_1 = L_2 = 0)}{P(L_1 = L_2)} = 1$, for:

$P(L_1 = L_2, L_1 \neq 0) = \int_{\mathbb{R} \setminus \{0\}} P(L_2 = l_1) d\mu_{L_1}(l_1) = 0$